\documentclass[proof]{WileyASNA-v1}

\articletype{Article Type}%
\usepackage{caption}
\usepackage{ amssymb }
\usepackage{ mathrsfs }
\usepackage{graphicx}

\received{xx November 2023}
\revised{xx November 2023}
\accepted{xx November 2023}

\newcommand{\old}[1]{{\color[rgb]{0.5,0.5,0.5}\sout{#1}}}

\begin{document}


\title{Differential Rotation in Compact Objects with Hyperons and Delta Isobars}

\author[1]{Delaney Farrell*}

\author[1,2]{Fridolin Weber}

\author[3]{Jia Jie Li}

\author[4,5]{Armen Sedrakian}

\authormark{Farrell \textsc{et al}}

\address[1]{Department of Physics, San Diego State University, San Diego, CA 92115, United States}

\address[2]{Center for Astrophysics and Space Sciences,
University of California at San Diego, San Diego, CA 92093, United States}

\address[3]{School of Physical Science and Technology, Southwest University, Chongqing 400715, China}

\address[4]{Frankfurt Institute for Advanced Studies, D-60438 Frankfurt am Main, Germany}

\address[5]{Institute of Theoretical Physics, University of Wroclaw, 50-204 Wroclaw, Poland}


\corres{*\email{dfarrell@sdsu.edu}}

\abstract{Neutron stars may experience differential rotation on short, dynamical timescales following extreme astrophysical events like binary neutron star mergers. In this work, the masses and radii of differentially rotating neutron star models are computed. We employ a set of equations of states for dense hypernuclear and $\Delta$-admixed-hypernuclear matter obtained within the framework of CDF theory in the relativistic Hartree-Fock (RHF) approximation. Results are shown for varying meson-$\Delta$ couplings, or equivalently the $\Delta$-potential in nuclear matter. A comparison of our results with those obtained for non-rotating stars shows that the maximum mass difference between differentially rotating and static stars is independent of the underlying particle composition of the star. We further find that the decrease in the radii and increase in the maximum masses of stellar models when $\Delta$-isobars are added to hyperonuclear matter (as initially observed for static and uniformly rotating stars) persist also in the case of differentially rotating neutron stars. 
} 

\keywords{Neutron Star, Differential Rotation, Relativistic Hartree-Fock, Covariant Density Function Theory }

\maketitle

\section{Introduction}

As the densest observed stellar objects, neutron stars provide a unique, naturally occurring laboratory for studying matter at extreme pressures and densities not reproducible by experiments. The matter within the core of a massive neutron star can reach densities up to an order of magnitude higher than nuclear saturation density. One physical mechanism for supporting massive neutron stars is rotation, as rotating neutron stars can sustain a higher rest mass than their non-rotating counterparts. While we assume the majority of rotating neutron stars are rotating uniformly, some neutron stars may form with considerable differential rotation following extreme astrophysical events like core-collapse supernovas and binary neutron star mergers~\citep{morrison2004effect}. The state of differential rotation is short-lived, as stars will relax to uniform rotation due to the shear viscosity of matter on timescales that are much shorter than secular timescales.

Studies of differentially rotating neutron stars are of particular interest in the context of binary neutron star mergers following the recent first observation of a binary neutron star merger, GW170817~\citep{LIGO-Virgo:2017}. Binary neutron star mergers 
are likely to form differentially rotating remnants as evidenced by general relativistic numerical simulations and expected from complicated hydrodynamic motions during the coalescence. 
A differentially rotating remnant star formed following a merger event has been confirmed through numerical simulation in the past \citep{shibata2017gravitational, shibata2000simulation, fujibayashi2018mass}. As we move into a new era of multi-messenger astronomy following the GW170817 event, understanding not only the inspiral phase but also the post-merger phase has the potential to provide further information on the equation of state of dense matter, neutron star properties, and the remnant's evolution.

Following a binary neutron star merger, the remnant star can take the shape of a hypermassive or supermassive neutron star \citep{sarin2021evolution, hotokezaka2013remnant}, or promptly collapse into a black hole. Differential rotation is one mechanism that allows these stars to sustain total masses considerably higher than both non-rotating stars and uniformly rotating stars. These massive, differentially rotating neutron stars may also deviate from spherical or axial symmetry by exhibiting extreme triaxial deformations. Both the structural deformation and differential rotation allow the compact remnant to remain stable on short, dynamical timescales, for masses that would render configurations unstable in the static and uniform-rotation cases.

The observable properties of a neutron star, like its mass, are not only dependent on rotation. Additionally, the underlying microphysics, given by the equation of state (EOS), greatly influences the overall structure of these objects. The cores of neutron stars can reach densities up to a few times nuclear saturation density, covering a density and temperature regime not reproducible in laboratories and not fully understood. 
The high-density regime is expected to not only contain nucleons, but also other exotic degrees of freedom like deconfined quark matter, hyperons, and delta isobars ($\Delta$). These additional degrees of freedom have been shown previously in numerical simulations to greatly impact properties like the mass and radius of neutron stars. In \cite{li2018competition}, the presence of $\Delta$'s was shown to soften the EOS in the low to intermediate-density region and stiffen it at high densities, resulting in a slightly increased maximum mass and considerably decreased radius for non-rotating compact objects.
In this work, we extend the work done in~\cite{li2018competition}
and \cite{Li:2023owg} to explore differential rotation's impact on both stars containing hyperonic matter and $\Delta$ populations. The paper is organized as follows: Section~\ref{background} describes the theoretical framework for modeling differential rotation in neutron stars and for constructing the dense matter EOS in covariant density functional (CDF) theory. Section~\ref{results} presents calculated mass-radius relations for differentially rotating stars containing both hyperon and $\Delta$ isobar populations, and density maps indicating at what depths different particle species appear within these stars. Section~\ref{discussion} gives a summary of the work presented.

\section{Background}\label{background}

\subsection{Modeling Differential Rotation in Neutron Stars}

To calculate the bulk properties of neutron stars, the stellar structure equations must be solved in the framework of Einstein's theory of general relativity. These equations are derived from Einstein's field equation and depend on the stellar matter's EOS, i.e., the underlying relationship between pressure $P$ and energy density $\epsilon$. In general, modeling uniformly or differentially rotating neutron stars is more complicated than modeling their non-rotating counterparts. Rapid rotation of either kind can deform the structure of the star, resulting in a flattening at the pole and expansion along the radius in the equatorial direction. To account for this deviation from spherical symmetry, the stellar structure equations for rotating stars must be dependent on both the radial coordinate $r$ and the polar coordinate $\theta$, assuming axisymmetry. Rotation also results in more massive objects, as it is a physical mechanism that stabilizes massive stars against collapse; as a result, rotating neutron stars are able to sustain roughly up to 20\% more mass than their non-rotating counterparts, depending on the underlying equation of state~\citep{kalogera1996maximum}. The increased mass of rotating stars alters the geometry of spacetime by introducing a dependence on the star's rotational frequency to the line element and a self-consistency condition to the stellar structure equations to account for the dragging of local inertial frames \citep{Weber:1999book}.   

Modeling differential rotation in neutron stars begins with the same metric as uniformly rotating stars, which is dependent on both the radial coordinate $r$ and polar angle $\theta$:
\begin{equation}
    ds^2 = -e^{\gamma-\rho} dt^2 +  e^{2\alpha}(dr^2 +r^2d\theta) + e^{\gamma-\rho}r^2 \sin^2\theta (d\phi - \omega dt)^2,
\end{equation}
where $\gamma$, $\rho$, $\alpha$, and $\omega$ are metric functions, where dragging of local inertial frames is accounted for by $\omega$. These functions are dependent on $r$ and $\theta$, as well as the star's angular velocity $\Omega$. The metric functions are computed from Einstein's field equation
\begin{equation}
    R^{\kappa \sigma} - \frac{1}{2} Rg^{\kappa \sigma} = 8\pi T^{\kappa \sigma},
\end{equation}
where $R^{\kappa \sigma}$ is the Ricci tensor, $R$ is the curvature scalar, and $g^{\kappa \sigma}$ is the metric tensor. The energy momentum tensor, $T^{\kappa \sigma}$, is given by
\begin{equation}
    T^{\kappa \sigma} = (\epsilon + P) \, u^\kappa u^\sigma + g^{\kappa \sigma} P,
\end{equation}
where $\epsilon$ and $P$ are given by the underlying EOS.

Once the four metric functions are solved for, they are used to solve the equation of hydrostatic equilibrium for a barotropic fluid:
\begin{equation}
    h(P) - h_p = \frac{1}{2} [\gamma_p + \rho_p - \gamma - \rho - \text{ln}(1-v^2) + F(\Omega)]
    \label{hydro}
\end{equation}
where $h(P)$ is the enthalpy, $\gamma_p$ and $\rho_p$ are the value of the metric potentials at the pole $p$, and $v$ is defined as
\begin{equation}
    v = (\Omega - \omega) \, r \, \sin \, \theta e^{-\rho}. 
\end{equation}
The final term in Eqn.~\ref{hydro}, $F(\Omega)$, defines the rotational law for the matter. Previous work \citep{komatsu1989rapidly,cook1992spin,cook1994rapidly} defines the rotation law as the linear function
\begin{equation}
    F(\Omega) = A^2 (\Omega_c - \Omega),
    \label{rotlaw}
\end{equation}
where $\Omega_c$ is the central value for the angular velocity. The parameter $A$ is used to determine the length scale over which the frequency changes, and acts as a scaling factor of the degree of differential rotation~\citep{morrison2004effect}. 

Substituting Eqn.~\ref{rotlaw} into Eqn.~\ref{hydro} gives 
\begin{equation}
\label{eq:7}
    (\Omega_c - \Omega) = \frac{1}{\hat{A}^2} \left[ \frac{(\Omega-\omega)\, s^2 (1-\mu^2) \,e^{-2\rho}}{(1-s)^2-(\Omega-\omega)^2\, s^2\, (1-\mu^2)\, e^{-2\rho}} \right],
\end{equation}
where the rotational frequency $\Omega$ can be then isolated and solved numerically using a root-finding algorithm like Newton-Rhapson. The parameters $s$ and $\mu$ in Eqn.~\eqref{eq:7} are dimensionless representations of the radial and polar coordinates. Additionally, Eqn.~\eqref{eq:7} uses a modified version of the rotation parameter, $\hat{A}$, which is scaled by the equatorial radius $r_e$: $\hat{A} \equiv A/r_e$. Uniform rotation is achieved in the limit $\hat{A}^{-1} \rightarrow 0$. Bulk properties of the star, like mass, angular momentum, and rotational kinetic energy, are calculated once a set of self-consistent solutions to the above equations is found.

\subsection{Equation of State}\label{back_eos}

The models of dense matter EOS  used in this work are based on the framework of CDF theory, where meson-$\Delta$ coupling values are varied and the calculations are carried out with the relativistic Hartree-Fock (RHF) approximation. The Lagrangian of the model is given as
\begin{equation}
    \mathscr{L} = \mathscr{L}_{\text{B}} + \mathscr{L}_{\text{m}} + \mathscr{L}_{\text{int}} + \mathscr{L}_{\text{l}},
\end{equation}\label{eq:lagrangian}
where $\mathscr{L}_{\text{B}}$ represents free baryonic fields $\psi_{\text{B}}$, $\mathscr{L}_{\text{m}}$ represents free meson fields $\phi_{\text{m}}$, $\mathscr{L}_{\text{int}}$ describes the interaction between baryons and mesons, and $\mathscr{L}_{\text{l}}$ represents the contribution from free leptons. The baryons accounted for include the spin-1/2 octet of nucleons $N \in \{n, p\}$ and hyperons $Y \in \{\Lambda, \Xi^{0, -}, \Sigma^{+, 0, -}\}$ and the spin-3/2 zero-strangeness quartet $\Delta \in \{\Delta^{++, +, 0, -}\}$. The mesons accounted for include those regularly encountered in the RHF approximation including the isoscalar-scalar meson $\sigma$, the isoscalar-vector meson $\omega$, the isovector-vector meson $\rho$, and the pseudo-vector meson $\pi$, as well as two hidden-strangeness mesons, $\sigma^*$ and $\phi$, which describe the interaction between hyperons. The leptons accounted for are electrons $e^-$ and muons $\mu^-$. 

The standard procedure for obtaining the density functional begins with finding equations of motion from the Euler-Lagrange equations for each particle species. These take the form of the Dirac equations for the baryon octet and leptons, the Rarita-Schwinger equations for the $\Delta$'s, and the Klein-Gordon equations for the mesons. The solutions of these equations are used to generate the energy density functional by evaluating each of the baryon self energies, $\Sigma$, in the RHF approximation. In Dirac space, the self-energy can be defined as
\begin{equation}
    \Sigma(k) = \Sigma_S(k) + \gamma_0 \Sigma_0(k) + \boldsymbol{\gamma} \cdot \boldsymbol{\hat{k}} \Sigma_V(k),
\end{equation}
where $\Sigma_S$, $\Sigma_0$, and $\Sigma_V$ denote the scalar self-energy, the time and space components of vector self-energy, respectively,  $\gamma_\mu$ ($\mu$ = 0-3) are the Dirac matrices, and $\boldsymbol{\hat{k}}$ is a unit vector along the 3-momentum $\boldsymbol{k}$. The self-energies, field equations, and the required charge neutrality condition are then used to determine EOS at zero temperature. The full self-consistent procedure is outlined in greater detail in~\cite{li2018competition, sedrakian2023heavy, Weber:1999book}.

\begin{figure}
    \centering
    \includegraphics[width=8.5cm]{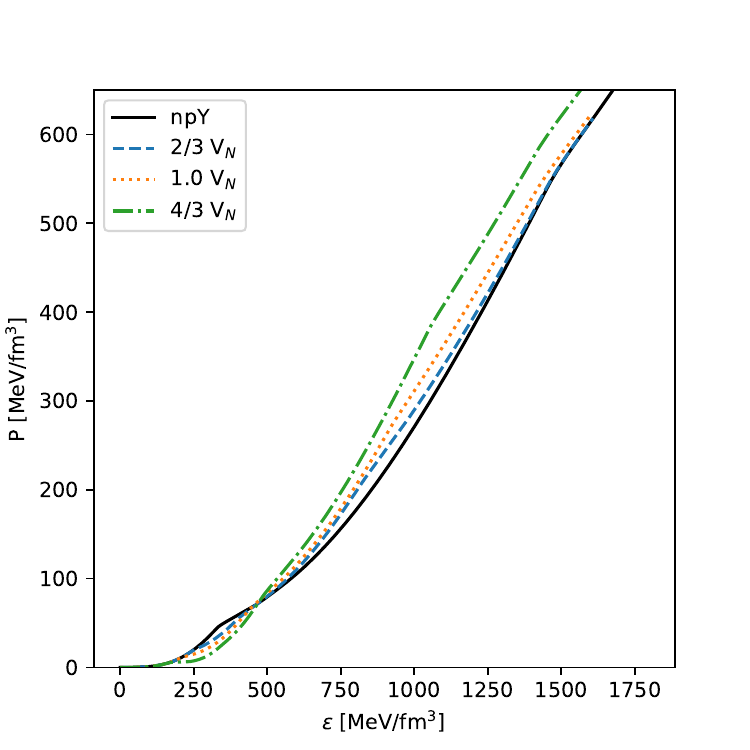}
    \caption{Pressure $P$ vs. energy density $\epsilon$ for four EOS models. The black curve, denoted "npY", describes hypernuclear matter that does not contain $\Delta$ isobars. The other three curves contain both hyperons and $\Delta$'s, parametrized by different values of the potential depth V$_\Delta$ (see text for more details). }
    \label{fig:eos_plot}
\end{figure}

The interaction between mesons and baryons, described by $\mathscr{L}_{\text{int}}$ in Eqn.~\eqref{eq:lagrangian}, is parametrized by meson-baryon coupling constants $g_{mB}$. The values for each of the coupling constants are dependent on the type of baryon and must be fitted to empirical data of nuclear and hypernuclear systems. For the hypernuclear sector, the coupling constant values are given by the SU(3) flavor symmetry quark model for vector mesons; for scalar mesons, the coupling constants are fitted to empirical hypernuclear potentials. The meson-$\Delta$ coupling is parametrized by $\Delta$ isoscalar potential V$_\Delta$, which is measured in units of the nucleon isoscalar potential V$_{\text{N}} = \Sigma_{0,\omega}(\omega) + \Sigma_{S,\sigma}(\sigma)$. Heavy-ion collision and scattering experiments provide approximate upper and lower bounds on V$_\Delta$ relating to V$_{\text{N}}$, but there are no exact values defined for the isoscalar potential. Following~\cite{li2018competition}, three potential depths V$_\Delta$ are chosen at nuclear saturation density $\rho_0$ as V$_\Delta$ = 2/3 V$_{\text{N}}$, V$_{\text{N}}$, and 4/3 V$_{\text{N}}$. These are consistent with inferences of $\Delta$ potentials from terrestrial experiments and are compatible with the lower limit on the maximum mass of static compact stars $M_{\rm TOV}\ge 2M_\odot$. For comparison, we also include a hypernuclear EOS constructed from the framework of CDF theory which does not include $\Delta$ particles, labeled "npY" throughout the text. The three mixed $\Delta$-hyperon EOS models parametrized by V$_\Delta$ and the purely hypernuclear EOS model npY are shown in Fig.~\ref{fig:eos_plot}. 

\section{Results}\label{results}

The EOS models described in Section~\ref{back_eos} were used to calculate the properties of differentially rotating neutron stars. As demonstrated in~\cite{li2018competition}, accounting for $\Delta$'s reduced radii of non-rotating stars when compared to the hypernuclear only EOS, npY. Additionally, including $\Delta$'s increased slightly the masses of non-rotating stellar sequences, where the larger by absolute value depth of the potential V$_\Delta$ resulted in a higher maximum mass. In the following sections, stellar sequences under different degrees of differential rotation are calculated to compare to the non-rotating mass and radius values shown in~\cite{li2018competition} to explore if the same trends are replicated. Furthermore, we calculate density maps of high-mass, differentially rotating stars to demonstrate at what depth specific particle species (hyperons and/or $\Delta$'s) appear. 

\subsection{Stellar Sequences}\label{sequences}

We first show results for stellar sequences with a constant central density range in order to demonstrate the expected increase in mass due to differential rotation. Because the rotation parameter repeatedly appears as $\hat{A}^{-1}$ in the equations described in Section~\ref{background}, we follow the lead of previous work which parameterized sequences by values of $\hat{A}^{-1}$ = 0.3, 0.5, 0.7, and 1.0~\citep{morrison2004effect, cook1992spin, galeazzi2012differentially}. As mentioned previously, uniform rotation is obtained in the limit $\hat{A}^{-1} \rightarrow 0$, and an upper bound of the scaled rotation parameter is $\hat{A}^{-1} = 1.0$. As differential rotation becomes more extreme in the star ($\hat{A}^{-1} \rightarrow 1.0$), the maximum mass and equatorial radius of a stellar sequence is expected to increase.  

\begin{figure}
    \centering
    \includegraphics[width=8.5cm]{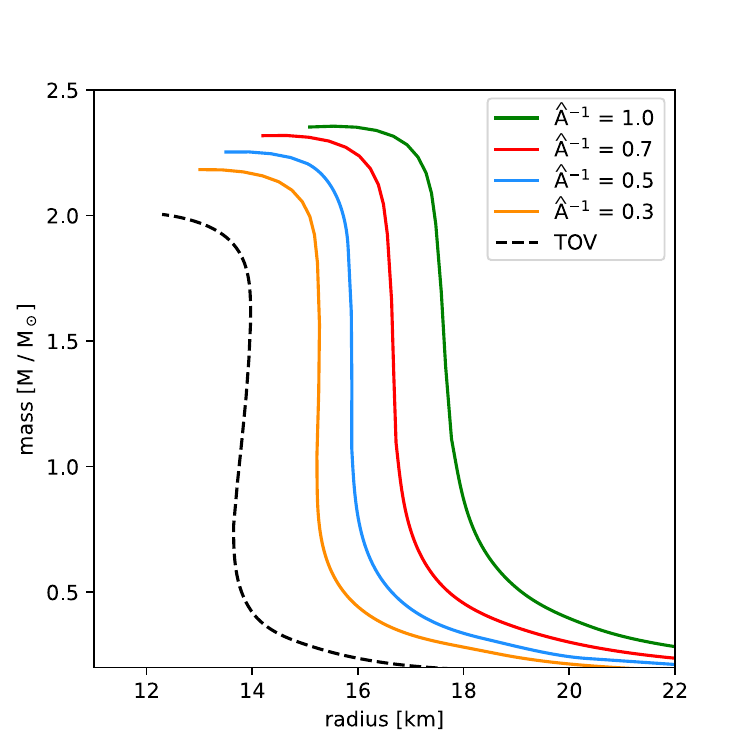}
    \caption{Mass-radius relations for the hyperonic EOS npY. The dashed black line shows the non-rotating curve resulting from the TOV equation, while the solid-colored lines are curves under different degrees of differential rotation. Differential rotation is parameterized by values of the rotation parameter, $\hat{A}^{-1}$ (see text for more details). }
    \label{fig:mr_npy}
\end{figure}

\begin{figure}
    \centering
    \includegraphics[width=8.5cm]{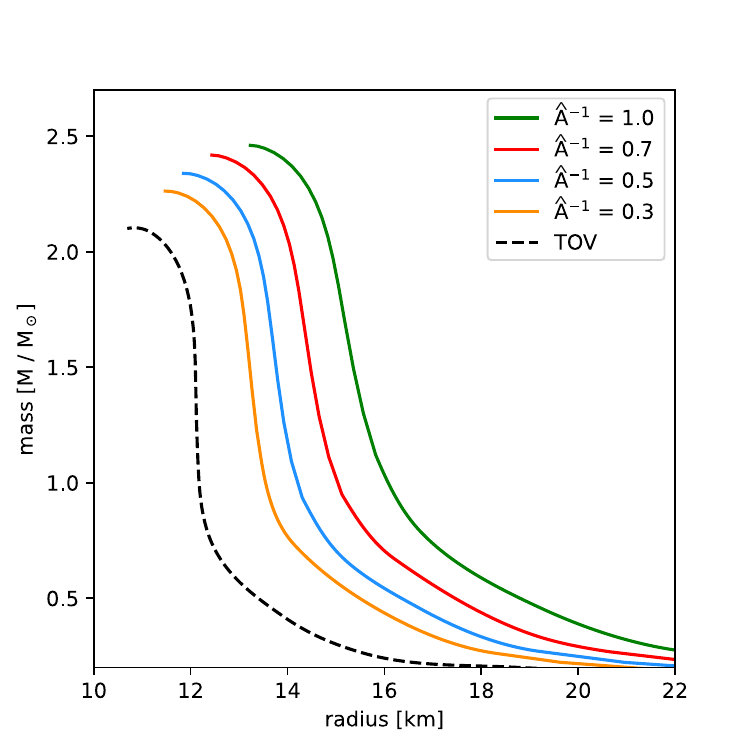}
    \caption{Mass-radius relations for the mixed hyperon-$\Delta$ EOS with V$_\Delta$ = 4/3 V$_{\text{N}}$. The dashed black line shows the non-rotating curve resulting from the TOV equation, while the solid-colored lines are curves under different degrees of differential rotation. Differential rotation is parameterized by values of the rotation parameter, $\hat{A}^{-1}$ (see text for more details).}
    \label{fig:mr_dc}
\end{figure}

This is indeed the case for the EOS containing hyperonic matter (npY) and the three EOS which vary $\Delta$ potential (V$_\Delta$) depths. Figures~\ref{fig:mr_npy} and \ref{fig:mr_dc} show mass-radius (equatorial) relations for four degrees of differential rotation, parameterized by $\hat{A}^{-1}$, compared to the non-rotating curves calculated with the Tolman-Oppenheimer-Volkoff (TOV) equation; Figure~\ref{fig:mr_npy} uses the hyperonic EOS model \old{denoted by npY,} and Figure~\ref{fig:mr_dc} uses the EOS model with largest potential depth value of V$_\Delta$ = 4/3 V$_{\text{N}}$~\cite{li2018competition}.

They highlight that the inclusion of $\Delta$ in the EOS composition \old{will} soften the EOS at lower densities and stiffens it at higher densities. This impacts the corresponding mass-radius relations for non-rotating stars by decreasing the radius but increasing the maximum mass. When comparing the mass-radius curves under differential rotation resulting from the hyperonic EOS in Figure~\ref{fig:mr_npy} to those resulting from the largest value of V$_\Delta$ in Figure~\ref{fig:mr_dc}, the same trend is observed. For the highest degree of differential rotation ($\hat{A}^{-1} = 1.0$), the canonical 1.4 M$_\odot$ star containing $\Delta$'s has a radius of 15.46 km for the EOS with V$_\Delta$ = 4/3 V$_{\text{N}}$, compared to a radius of 17.65, reflecting the same decrease in radius observed with the non-rotating curves in~\cite{li2018competition}.

The maximum masses of the hyperonic EOS and the three EOS models containing both hyperons and $\Delta$'s are given in Table~\ref{tab:maxmass} for non-rotating (TOV) and differentially rotating stars. For values of parameter  $\hat{A}^{-1}$ corresponding to different degrees of differential rotation are used. Both Table~\ref{tab:maxmass} and Figures~\ref{fig:mr_npy} and \ref{fig:mr_dc} reflect calculated values for stellar models at a fixed value of the ratio between the polar and equatorial radius, r$_{\text{ratio}}=0.7$. 
For all employed EOS models, we observe a similar 16-17\%  increase in the maximum mass of a maximally differentially rotating star when compared to the maximum mass of a static (TOV) star. This shows that the difference in particle compositions that were accounted for has no significant impact on how much the maximum mass will increase as the rotation parameter $\hat{A}^{-1} \rightarrow 1.0$. 

\begin{table*}[]
\centering
\caption{Maximum gravitational mass (in units of M$_\odot$) of four RHF EOS models which account for hyperons (npY) or hyperons and $\Delta$ particles (given as values of the V$_\Delta$ potential):}
\begin{tabular}{l|l|l|l|l|l}
\hline
  \textbf{EOS Model} & TOV & $\hat{A}^{-1}$ = 0.3 & $\hat{A}^{-1}$ = 0.5 & $\hat{A}^{-1}$  = 0.7 & $\hat{A}^{-1}$ = 1.0 \\ \hline
  npY              & 2.011 & 2.183 & 2.253 & 2.319 & 2.356 \\
  2/3 V$_\text{N}$ & 2.034 & 2.200 & 2.270 & 2.334 & 2.366 \\
  1.0 V$_\text{N}$ & 2.054 & 2.216 & 2.287 & 2.355 & 2.386 \\
  4/3 V$_\text{N}$ & 2.103 & 2.262 & 2.339 & 2.418 & 2.461 \\
\hline
\end{tabular}\label{tab:maxmass}
\end{table*}

\subsection{Particle Composition}\label{comp}

In this section, density distributions of massive, differentially rotating stars are used to demonstrate where in the star different particle species appear. 

Individual stellar models are constructed with a series of set parameters: the central density, the ratio between the polar and equatorial radius r$_{\text{ratio}}$, and the rotation parameter $\hat{A}^{-1}$ which specifies the degree of differential rotation. The parameter r$_{\text{ratio}}$ reflects the structural deformation observed with rapid rotation, which has been extensively showed through numerical simulation \citep{hartle1968slowly, huber1998neutron, Weber:1999book} to result in a lengthening of the radius at the equator and shortening at the pole. In the case of differential rotation, more extreme structural deformations can occur as r$_{\text{ratio}} \rightarrow 0$, in some cases leading to the formation of toroidally shaped objects \citep{lyford2003effects, morrison2004effect}. In this section, we fix r$_{\text{ratio}}$ at a value of 0.6, slightly lower than that used in Section~\ref{sequences}, resulting in massive, rapidly rotating, ellipsoid-shaped stars. 

\begin{figure}[ht]
\centering
\begin{minipage}[b]{.98\linewidth}
\centering
\includegraphics[width=.93\linewidth]{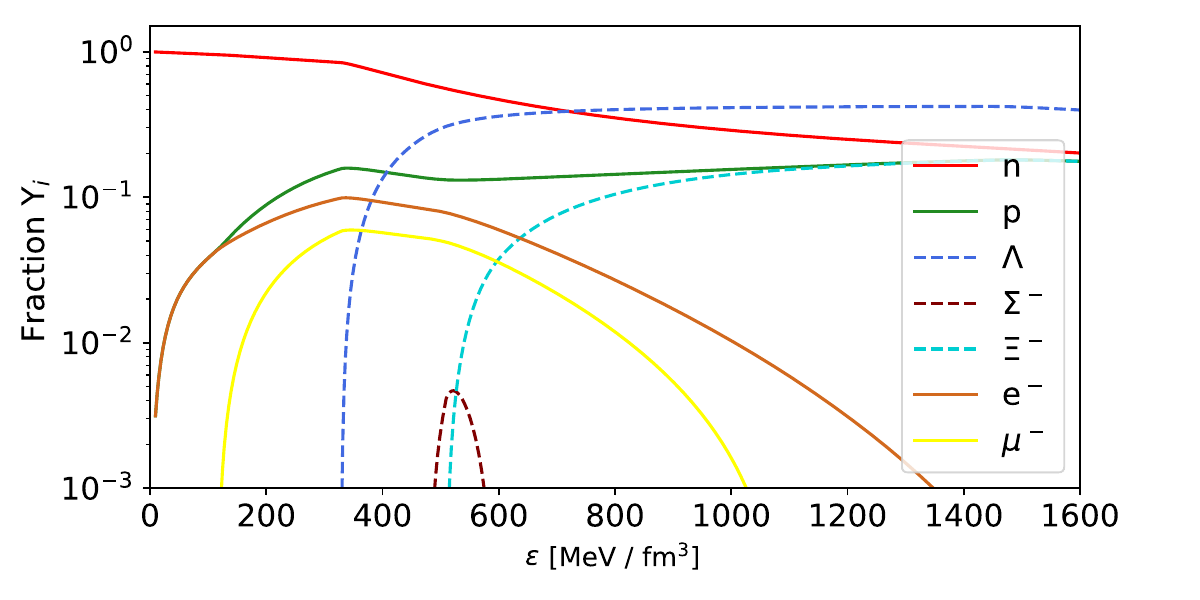}
\end{minipage}
\hfill
\begin{minipage}[b]{.98\linewidth}
\centering
\includegraphics[width=.95\linewidth]{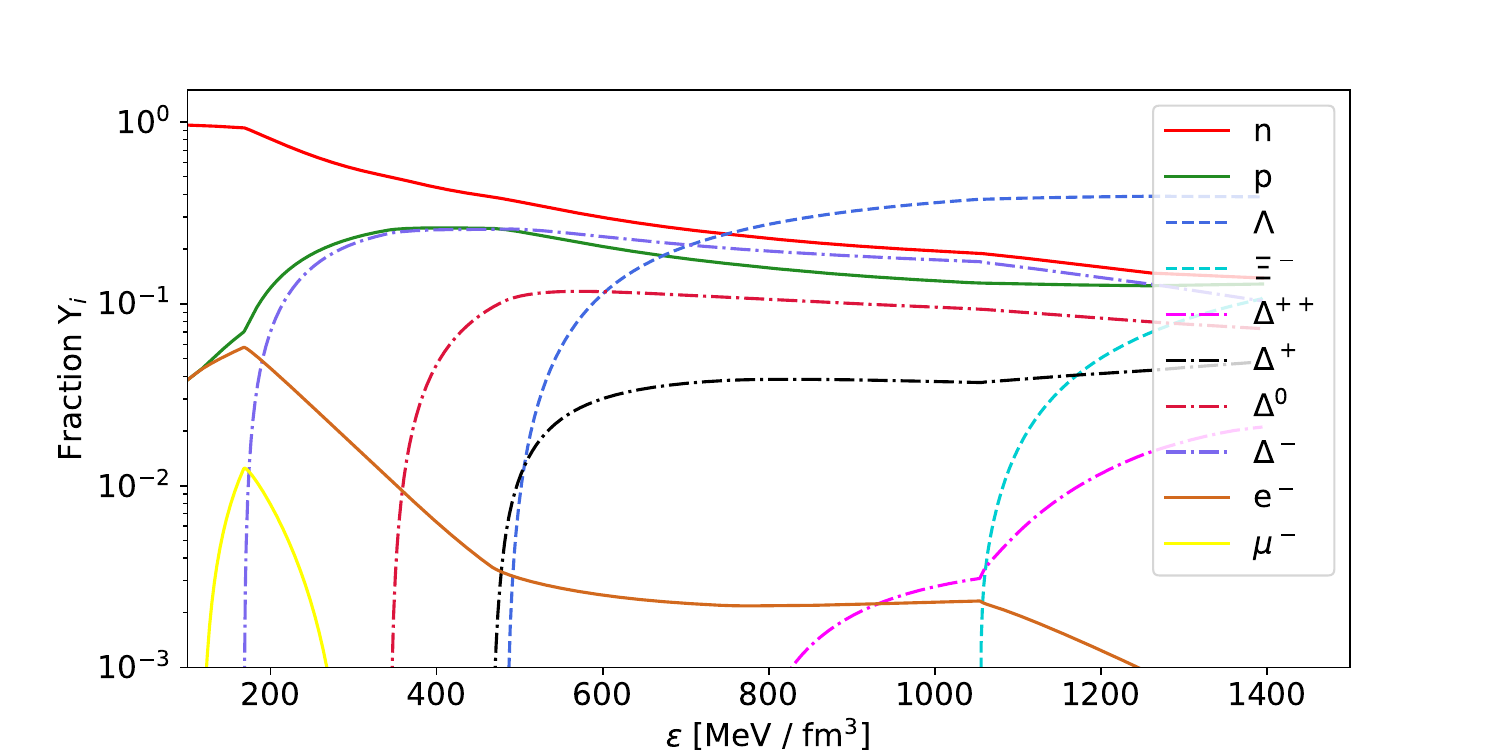}
\end{minipage}
\caption{Relative particle fractions Y$_i$ for the hypernuclear EOS npY (top panel) and the hyperon-$\Delta$ admixed EOS with a potential depth of 4/3 V$_{\text{N}}$ (bottom panel).  } 
\label{fig:particle_frac}
\end{figure}

\begin{figure}
    \centering
    \includegraphics[width=8.5cm]{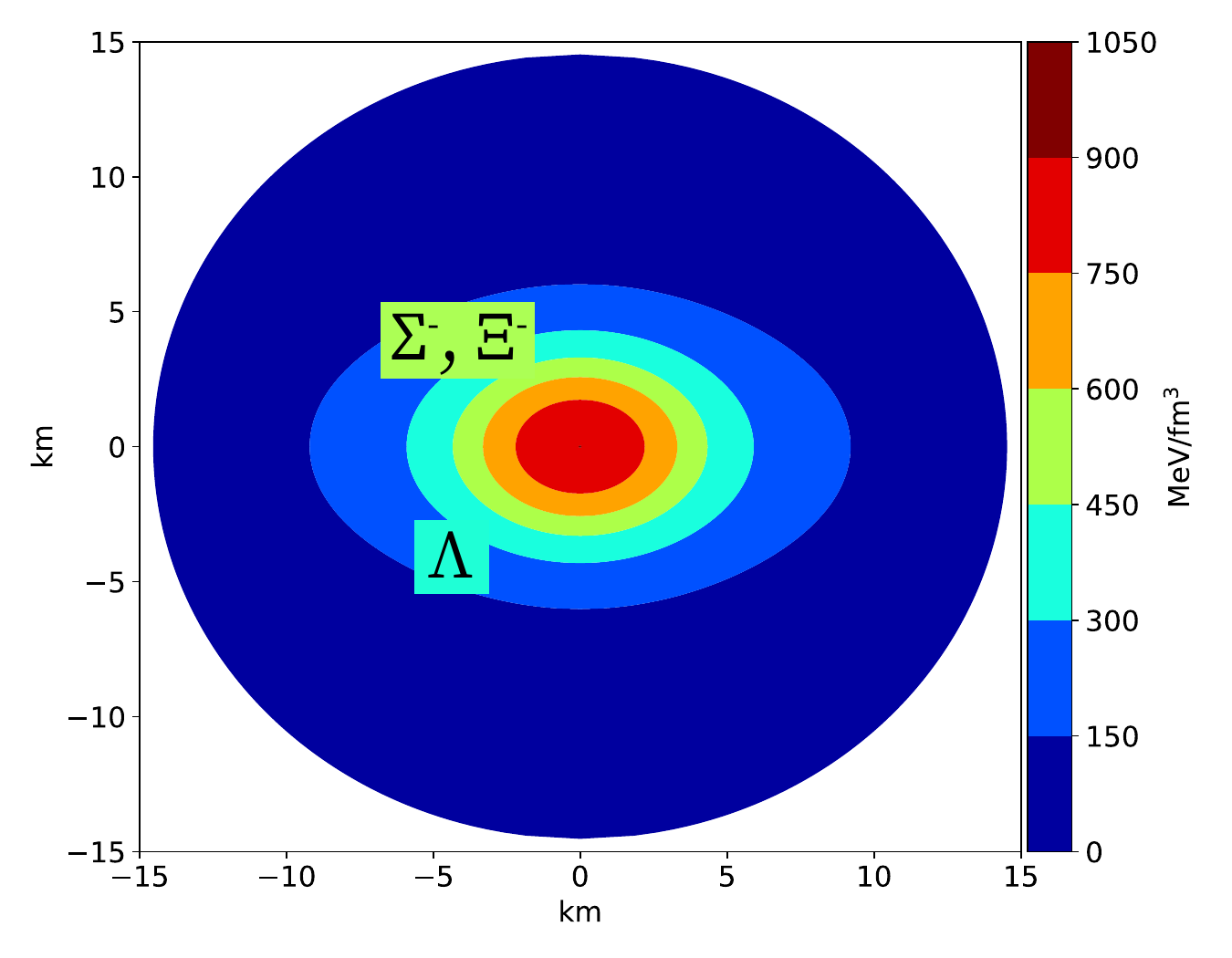}
    \caption{Energy density map of a 2.48 M$_\odot$ differentially rotating star constructed with hyperonic EOS npY. The star has an r$_\text{ratio}$ = 0.6 and rotation parameter $\hat{A}^{-1}$ = 0.7.  }
    \label{fig:single_hyperon}
\end{figure}

As in Section~\ref{sequences}, we construct density distribution maps for the hyperonic EOS model and the V$_\Delta$ = 4/3 V$_{\text{N}}$ EOS model additionally containing $\Delta$'s. Particle fractions for both EOS models can be found in Figure~\ref{fig:particle_frac}. For the hyperonic EOS model, the three hyperon species accounted for are the $\Lambda$, the $\Xi^-$, and the $\Sigma^-$ hyperons, which appear in that order. Figure~\ref{fig:single_hyperon} shows the density map of a 2.48 M$_\odot$ star constructed from the npY EOS model, with the following set parameters: central density $\epsilon_c$ = 900 MeV, r$_{\text{ratio}}$ = 0.6, and rotation parameter $\hat{A}^{-1}$ = 0.7. The $\Lambda$ hyperon first appears at an equatorial radius, r$_{\text{e}}$, of $\sim$5 km and a density of 329.4 MeV/fm$^3$ (shown in cyan); the $\Lambda$ hyperon would be present from this depth until the center of the star. The $\Sigma^-$ and $\Xi^-$ both appear at r$_{\text{e}} \approx$ 3 km and densities of 479.1 and 510.9 MeV/fm$^3$, respectively (shown in green). The $\Sigma^-$ only appear briefly due to their repulsive potential at nuclear saturation density \citep{li2018competition}, but the $\Xi^-$ would be populated throughout the star up to its center. 

\begin{figure}
    \centering
    \includegraphics[width=8.5cm]{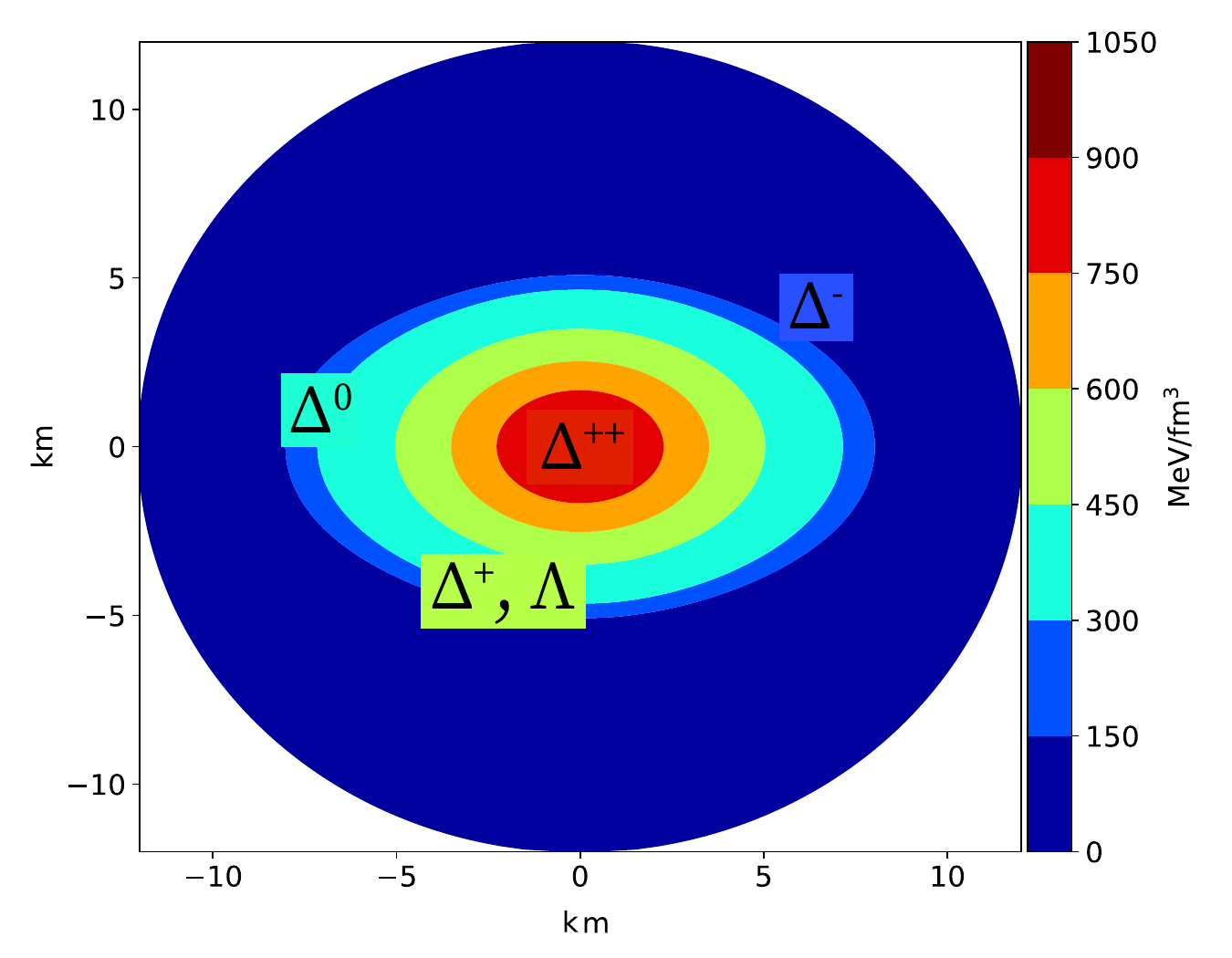}
    \caption{Energy density map of a 2.55 M$_\odot$ differentially rotating star constructed with hyperon-$\Delta$ admixed EOS with a potential depth of 4/3 V$_{\text{N}}$. The star has an r$_\text{ratio}$ = 0.6 and rotation parameter $\hat{A}^{-1}$ = 0.7.  }
    \label{fig:single_delta}
\end{figure}

The density distribution map for a 2.55 M$_\odot$ star constructed with the mixed hyperon and $\Delta$ EOS with a potential depth of V$_\Delta$ = 4/3 V$_{\text{N}}$ is shown in Figure~\ref{fig:single_delta}. The same values for the central density, r$_{\text{ratio}}$, and rotation parameter $\hat{A}^{-1}$ were used as in Figure~\ref{fig:single_hyperon}. As shown in both Figure~\ref{fig:particle_frac} and Figure 2 of~\cite{li2018competition}, a larger value of the potential V$_\Delta$ results in a lowered onset threshold for $\Delta$'s, and their threshold density is much lower than the one for the first hyperon. For the largest potential depth  V$_\Delta$ = 4/3 V$_{\text{N}}$, the entire spin-3/2 zero-strangeness quartet appears in the star. In this case, only the $\Lambda$ and $\Xi^-$ hyperons appear. The $\Sigma^-$ hyperon is effectively replaced by the same-charge $\Delta^-$ isobar, as it is energetically more favorable due to its large negative potential. In Figure~\ref{fig:single_delta}, the $\Delta^-$ appears at a density of 168.66 MeV/fm$^3$, corresponding to an r$_{\text{e}}$ of $\sim$8 km (shown in royal blue). The next species to appear, $\Delta^0$, appears at a density of 344.1 MeV/fm$^3$ corresponding to an r$_{\text{e}}$ of $\sim$8 km (shown in cyan). The $\Lambda$ and $\Delta^+$ become present at densities of 465.9 and 486.3 MeV/fm$^3$ respectively, shown in green at a radius of $\sim$ 5 km. The particle species to appear is the $\Delta^{++}$, at a density of 763.6 MeV/fm$^3$ and radius of $<$ 3 km shown in red. All particles would be present from the density they appear through to the center of the star. 

It is important to note here that neutrons, protons, electrons, and muons are additionally present within the stellar models discussed above but not shown explicitly in Figures~\ref{fig:single_hyperon} and \ref{fig:single_delta}.

\section{Discussion}\label{discussion}

In this work, we extended the previous work of ~\cite{li2018competition}
and \cite{Li:2023owg} that studied static and uniformly rotating stellar configurations of hyperonic and $\Delta$-admixed stars to stellar models that support differential rotation. Specifically, we examined 
how the inclusion of $\Delta$ isobars in the EOS of dense matter impacts the corresponding stellar properties like mass and equatorial radius for non-vanishing differential rotations. In Section~\ref{sequences}, 
we demonstrated that the inclusion of $\Delta$'s in hypernuclear EOS results in the reduction of the equatorial radius of the differentially rotating star and an increase in its maximum mass. These trends are consistent with analogous findings for uniformly rotating and static compact stars. An interesting finding is that when comparing the maximum masses from the highest degree of differential rotation (where the rotation parameter $\hat{A}^{-1} = 1.0$) to the lowest ($\hat{A}^{-1} = 0.3$), the magnitude of the increase in mass remained constant for all four EOS models considered. In particular, the inclusion of $\Delta$’s in the hypernuclear EOS does not change the total amount of mass increase once the differential rotation is allowed. In Section 3.2, density distribution maps of massive,
differentially rotating stars were shown to demonstrate at what radial depth different particle species appear. 
The competition between the nucleation of $\Sigma^-$ hyperon and $\Delta^-$ isobar was demonstrated, whose outcome depends on the value of the $\Delta^-$ potential in nuclear matter.
 
\section{Acknowledgements}
DF and FW are supported by the National Science Foundation (USA) under Grant No. PHY-2012152. J.~L. acknowledges the support of the National Natural Science Foundation 
of China (Grant No. 12105232), the Fundamental Research Funds for the Central 
Universities (Grant No. SWU-020021), and by the Venture \& Innovation Support 
Program for Chongqing Overseas Returnees (Grant No. CX2021007).
A.~S. is supported by the Deutsche Forschungsgemeinschaft Grant 
No. SE 1836/5-2 and the Polish NCN Grant
No. 2020/37/B/ST9/01937 at Wroclaw University.

\bibliography{ZenB}

\end{document}